\documentclass[journal=ancac3,manuscript=article]{achemso} 

\setkeys{acs}{maxauthors = 0} 

\usepackage[bookmarks=false,colorlinks=true, linkcolor=blue,urlcolor=blue,citecolor=blue, pdfstartview={FitH}]{hyperref}

\usepackage[utf8]{inputenc}
\usepackage{verbatim}
\usepackage{graphicx}
\usepackage{dcolumn}
\usepackage{bm}
\usepackage{float}
\usepackage{amsmath}
\usepackage{ulem}
\usepackage{xcolor}

\graphicspath{ {./FIG/} }

\newcommand{\onlinecite}[1]{\hspace{-1 ex} \nocite{#1} \citenum{#1}}
\newcommand{\rp}[1]{\textcolor{black}{#1}}
\newcommand{\rs}[1]{\textcolor{black}{#1}}
\newcommand{\rsf}[1]{\textcolor{black}{#1}}

\newcommand{\uamftmc}{Departamento de F\'{i}sica Te\'{o}rica de la Materia Condensada, Universidad Aut\'{o}noma de Madrid, E-28049 Madrid, Spain}
\newcommand{\ifimac}{Condensed Matter Physics Center (IFIMAC), Universidad Aut\'{o}noma de Madrid, E-28049 Madrid, Spain}
\newcommand{\julich}{Peter Gr\"{u}nberg Institut (PGI-3), Forschungszentrum J\"{u}lich, 52425, J\"{u}lich, Germany}
\newcommand{\jara}{J\"{u}lich Aachen Research Alliance (JARA), Fundamentals of Future Information Technology, J\"{u}lich, Germany.}
\newcommand{\expIIAachen}{Experimentalphysik II B, RWTH Aachen University; 52074 Aachen, Germany}
\newcommand{\expIVAachen}{Experimentalphysik IV A, RWTH Aachen University; 52074 Aachen, Germany}
\newcommand{\cologne}{Faculty of Mathematics and Natural Sciences, Institute of Physics II, University of Cologne, Cologne, Germany}
\newcommand{\inp}{Grenoble INP--Phelma, 3 Parv. Louis N\'{e}el, 38000 Grenoble, France}

\author{Emiliano Ventura-Macias} \affiliation{\uamftmc}
\author{Jose Martinez-Castro} \affiliation{\julich}  \alsoaffiliation{\jara}
\author{Guillermo Haas} \affiliation{\uamftmc} \alsoaffiliation{\inp}
\author{Jara Trujillo-Mulero} \affiliation{\uamftmc}
\author{Pablo Pou} \affiliation{\uamftmc} \alsoaffiliation{\ifimac}
\author{Taner Esat} \affiliation{\julich}  \alsoaffiliation{\jara}
\author{Markus Ternes} \affiliation{\julich} \alsoaffiliation{\jara} \alsoaffiliation{\expIIAachen}
\author{Ruslan Temirov} \affiliation{\julich} \alsoaffiliation{\jara} \alsoaffiliation{\cologne}  \email{r.temirov@fz-juelich.de}
\author{F. Stefan Tautz} \affiliation{\julich} \alsoaffiliation{\jara} \alsoaffiliation{\expIVAachen}
\author{Rub\'{e}n P\'{e}rez} \affiliation{\uamftmc} \alsoaffiliation{\ifimac} \email{ruben.perez@uam.es}

\title{Bond-resolved STM with density-based methods} 

\begin{document}

\begin{center}
\end{center}

\newpage    

\baselineskip24pt 

\begin{abstract}
Bond-resolved scanning tunneling microscopy (BRSTM) provides intramolecular contrast with CO-functionalized tips, but its interpretation requires a description of both probe relaxation and orbital-dependent tunneling. We introduce
a simulation framework that combines the Full-Density-Based Model (FDBM) for CO-tip relaxation with Chen's derivative rules applied directly to real-space density-functional-theory (DFT) wave functions. 
We test the method against a series of constant-height BRSTM images of perylene-3,4,9,10-tetracarboxylic dianhydride (PTCDA) on Ag(111), acquired in 10~pm tip-height increments, and published data for 1,5,9-trioxo-13-azatriangulene (TOAT) on Cu(111). The simulations reproduce the principal intramolecular contrast features and their evolution with tip height. The comparison identifies distinct roles for the substrate: it modifies the partially occupied molecular states and probe--sample forces in PTCDA/Ag(111), while its influence on the TOAT/Cu(111) contrast remains relevant even without substantial changes in molecular-orbital occupation. Our framework thus provides a practical and accurate route to interpreting BRSTM images that incorporates, from a single DFT calculation, the separate contributions of electronic structure and CO-tip relaxation.

\end{abstract}

{\textbf{Keywords:}} scanning tunneling microscopy, non--contact atomic force microscopy, tip functionalization, CO molecule, PTCDA, TOAT, DFT 

\newpage


Scanning tunneling microscopy (STM) is routinely employed to access and study the electronic properties of adsorbed molecules on metallic and insulating substrates like a supported NaCl bilayer~\cite{Pascual2000, Soe2009, Repp2006}.
Moreover, precise functionalization of the tip apex has allowed observation of their chemical structure~\cite{Weiss2010, Kichin2011} and frontier molecular orbitals~\cite{SoeACSNano2012, Villagomez2007, ReppPRL2005-2}.
Among the different possibilities for tip functionalization, carbon monoxide (CO) has been particularly useful due to its stability and capacity to resolve sub-molecular features.
These advantages were first exploited in frequency-modulation atomic force microscopy (FM-AFM) to achieve atomic resolution in organic molecules~\cite{gross2009}.
However, Bond-resolved STM (BRSTM)~\cite{SongBRSTMRev2021} can be more convenient experimentally as it does not require additional instrumentation for frequency-modulation~\cite{Nguyen2017BRSTMGNR}.
As such, it has been on the rise for achieving high resolution of various samples~\cite{Wang2024}, including single molecules~\cite{gross2011b, Song2020}, graphene nanorings and nanoribbons~\cite{Nguyen2017BRSTMGNR, Moreno2018, Lawrence2020, Li2021, Martinez-Castro2022}, and supramolecular assemblies where it supports the identification of halogen bonding~\cite{Lawrence2020a}.
\rs{In the case of purely carbon-based 2D materials, as nanographenes, letting aside some important electronic effects, images more or less resemble the network of C-C bonds (although with significant distortions). However, as soon as other chemical species are considered, the interpretation of BRSTM images becomes extremely  challenging, even for planar systems.
The  adsorption of 1,5,9-trioxo-13-azatriangulene (TOAT) on Cu(111) system is a good example, where the replacement of the central C atom with a N atom and of three H atoms by O atoms in triangulene leads to a very complex, distance-dependent, BRSTM contrast~\cite{HeijdenACSNano2016}.}
In light of the experimental advantages \rs{and interpretation challenges,} simulation models for BRSTM are highly desirable.

Sub-molecular resolution in BRSTM and the associated capability to visualize molecular structures and bonding topologies emerge when using a CO-functionalized tip at close range.
There, Pauli repulsion dominates due to the CO closed-shell electronic configuration~\cite{gross2009, Moll2010}.
This repulsion drives the CO away from areas with high charge density, i.e., from sample atoms and their chemical bonds.
Due to the tunneling current high sensitivity to distance, the CO deflection along bonds generates steep changes in contrast.
These changes appear as sharp edges in STM images, which gives its name to \textit{bond-resolved} STM (BRSTM).
On that account, the two key ingredients for any BRSTM model are CO deflection and tunneling through an adsorbed CO.
These two parts require the calculation of atomic forces acting in the junction 
to obtain the probe position (the oxygen position for a CO-tip) and an STM model to estimate the tunneling at that position.

\rs{So far, the PP-STM model~\cite{Krejci2017}  is the only approach proposed to simulate BRSTM images. It uses the Probe-Particle Model (PPM)~\cite{HapalaPRB2014} to  determine the CO deflection. In its early formulation, the method described probe-sample interactions using Lennard--Jones (LJ) potentials and accounted for CO bending through a harmonic term. Subsequent developments incorporated additional physical ingredients, such as electrostatic contributions modeled via point charges~\cite{HapalaElectricField2016}, and later extensions introduced electrostatics from the DFT sample charge density and other interaction terms fitted directly with DFT calculations~\cite{Oinonen2024}.}

\rs{ For the STM calculation, the PP-STM approach~\cite{Krejci2017} follows Bardeen's perturbative tunneling formalism, in which the tunneling matrix elements $M_{TS}$ are constructed by combining localized tip orbitals \rsf{(represented by symmetry-adapted $s$-like and $p$-like atomic-like orbitals)} with a local-orbital expansion of the sample states. In practice, \rsf{each $M_{TS}$ is calculated as a sum of tunneling matrix elements between a given tip orbital and local orbitals centered at the positions of each atom of the sample. These atom-centered orbitals are constructed via  a process that starts with a projection of the sample wave functions (obtained, in principle,  from a DFT calculation of the whole system, but, in most applications so far, just from the adsorbed molecule) onto localized atomic orbitals that are truncated spatially. 
Then,} the decay into vacuum is reconstructed using an effective work function and assumed exponential decay for each orbital. This approximation requires assigning a common decay parameter (via a work function) across orbitals and sites. Chen's derivative rules~\cite{Chen1990, Chen1998}  are then applied to generate the orbital-resolved contributions to $M_{TS}$, which are accumulated over all relevant tip--sample orbital pairs and over all atoms of the sample.}

\rs{While this approach is computationally efficient and often reasonable for some molecular states, it becomes much more delicate when mixing orbitals of fundamentally different character, such as molecular states and metallic substrate states. In such cases, the decay behavior is not uniform, and the use of truncated bases combined with ad hoc decay reconstruction complicates a quantitatively consistent treatment of the substrate. 
%
%
Moreover, the atom-by-atom construction of the tunneling matrix elements naturally scales with the number of atoms considered. As a result, the method is very efficient when only the atoms of the adsorbed molecule are included --while treating its coupling to the substrate effectively, but becomes more computationally demanding when the explicit substrate is incorporated.}

\begin{figure}[p]
	\centering
	\includegraphics[width=1.00\textwidth]{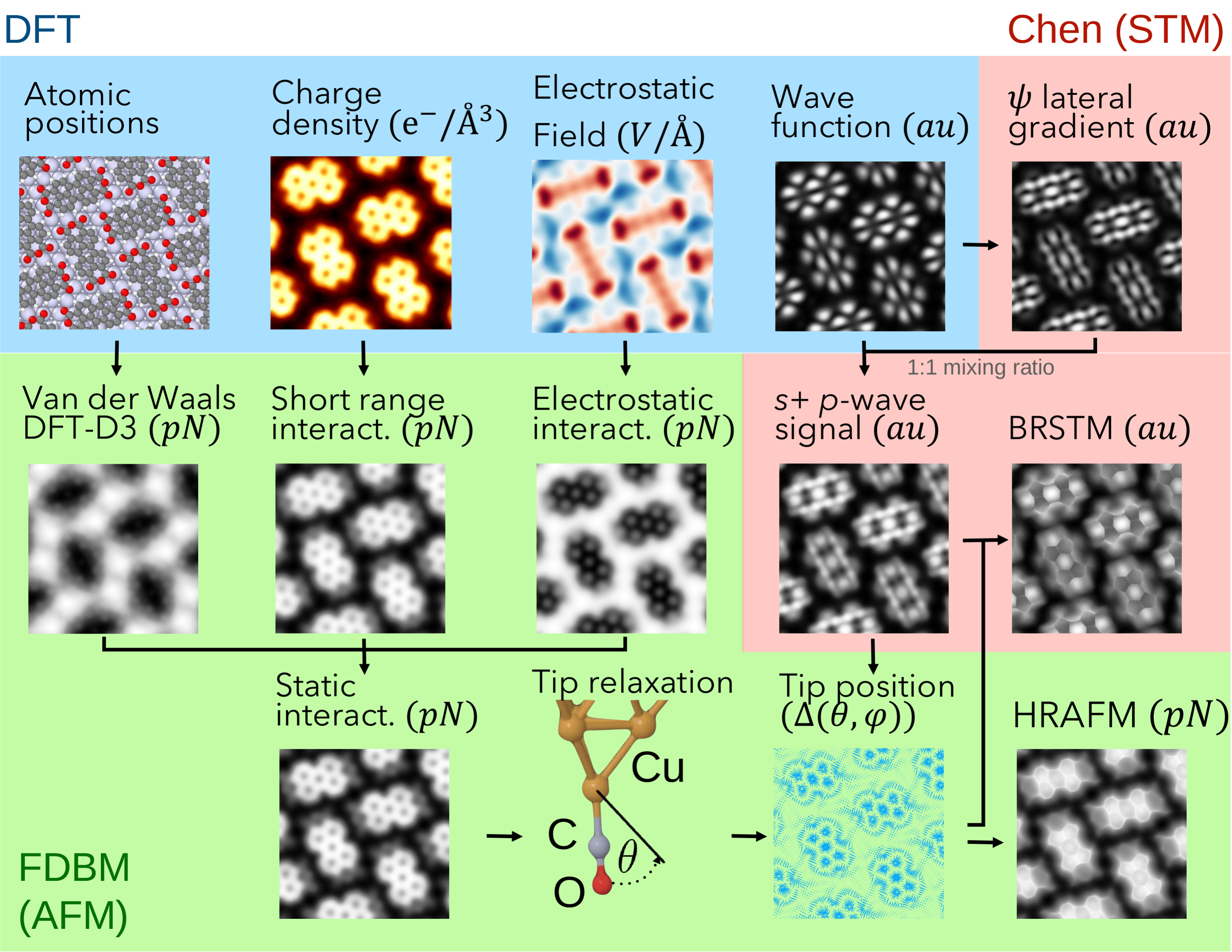}
        \caption{Simulation framework to obtain BRSTM (red area) and HRAFM (green area) images from a single DFT calculation (blue area). The atomic positions, charge density, and electrostatic potential of the sample are fed into the vdW, SR, and ES calculations, respectively. For the last two, the tip's charge density is also included. The three calculated interactions are added together to obtain the static force. Then, the tip relaxation is performed following a tilting constraint, from which the CO deflection and the HRAFM signal (the force, as shown here, 
\rsf{or, the frequency shift,  through a proper integration of the force~\cite{EVMAppSurfSci2023}) are obtained.}
\rsf{For the tunneling part, the CO 5$\sigma$  and 2$\pi^*$ orbitals are represented by an spherical \textit{s}-like orbital and two \textit{p}-like orbitals with its lobes pointing in the $x$ and $y$ cartesian axes in a plane parallel to the surface, respectively.}
Following Chen's approximation~\cite{Chen1990, Chen1998, gross2011b}, the squared modulus of  the DFT wave function is used to obtain the $s$-tunneling component, and the square modulus of its lateral gradient is calculated to get the $p$ component. These two components are added and integrated over the applied bias to get the $s+p$ signal. Lastly, the probe position (the position of the O atom once the CO deflection has been determined from the HRAFM calculation) is used to obtain the approximate tunneling for each tip position, resulting in the BRSTM signal. 
}
        \label{fig:framework}
\end{figure}

\rs{Here, we propose a practical alternative for BRSTM simulations that operates directly on real-space wavefunctions obtained from \rsf{a DFT calculation of the sample (molecule+substrate)}, thereby naturally incorporating the full spatial decay behavior of both molecular and substrate states without additional assumptions. Our approach  uses Chen's derivative rule~\cite{Chen1990, Chen1998} directly on the DFT-calculated sample's wave functions~\cite{gross2011b,Paschke2025} to include the tunneling from the 5$\sigma$ and 2$\pi^*$ CO orbitals (the HOMO and LUMO for the isolated CO molecule), as they both contribute due to their interaction with the electronic states of the metal tip~\cite{Drakova2006, Paulsson2008, Tiwari2009}. This fast approach, that delivers the proportional conductance at each point of the DFT-defined grid, is combined with the Full Density-Based Model (FDBM)~\cite{ellner2019, EVMAppSurfSci2023} to take into account the CO deflection  induced by the interaction with the sample.}

\rs{FDBM is designed to reproduce tip--sample forces obtained from DFT by summing up three force contributions calculated using the electron densities and electrostatic potentials of the isolated tip and sample as input, as illustrated in Figure~\ref{fig:framework}. The force components of FDBM are represented in the green area of Figure~\ref{fig:framework}, and their respective DFT inputs in the blue area. The short-range Pauli repulsion (SR) is calculated from the overlap of the probe and sample's charge densities ($\rho$)~\cite{ellner2019}.
Then, the electrostatic interaction (ES) is directly obtained from the sample's electrostatic potential ($\phi$) and the tip's $\rho$. Including the sample's $\rho$ and $\phi$ is what allows FDBM to reproduce HRAFM  features associated with charge redistribution~\cite{zahl2021TMA}.
van der Waals  (vdW) contributions are described using the widely employed DFT-D3 approach~\cite{Grimme2010}. DFT-D3 takes into account the chemical environment of an atom in order to determine the associated C$_{6}$ coefficient for the vdW calculation, improving significantly the accuracy as shown by  tests against high-level quantum chemistry calculations.
Finally, the CO deflection is 
described using a model of a lever with a torsional constraint on the polar angle.
For the equations that describe the SR, ES, vdW, and CO deflection, see Methods 
and Refs.~\onlinecite{ellner2019,EVMAppSurfSci2023}.
The real probe position (the position of the O atom once the CO deflection has been considered) is efficiently determined by minimizing the total energy  using a 3D-periodic tri--cubic interpolator~\cite{Lekien2005}.}

\rs{In our approach, the 5$\sigma$ orbital is approximated as an spherical \textit{s}-like orbital and the 2$\pi^*$ as two \textit{p}-like orbitals with its lobes pointing in the $x$ and $y$ cartesian axes in a plane parallel to the surface~\cite{Chen1990, Chen1998, gross2011b}.
While Tersoff-Hamann theory can be used to model the {s}-like orbital~\cite{TersoffHamann}, Chen's approach is needed for the \textit{p}-like orbitals~\cite{Chen1990, Chen1998}.
In short, this is the squared modulus of the lateral gradient of the sample wave function (see Methods, Eq.~\ref{eq:pwave})---while the \textit{s}-orbital is just the squared modulus (Methods, Eq.~\ref{eq:swave}), as illustrated in the pink area of Figure~\ref{fig:framework}.
The CO deflection calculated for each tip position with FDBM determines the O atomic coordinates (from now on referred as the probe position) where we evaluate both contributions to determine the total  tunneling current.}

\rs{Thus, our BRSTM model allows for a complete self-contained framework to simulate HRAFM and BRSTM images from a single DFT calculation. Importantly, because the method requires only the real-space representation of the sample wave function, it is fully compatible with any underlying DFT basis set --plane waves, LCAO, or others. Furthermore, it naturally incorporates, with no added computational cost,  both the structural and electronic effects of the substrate: the first one is  already accounted for when performing the DFT geometry optimization of the adsorbed molecule, while the second emerges naturally with the use of the sample (molecule + substrate) functions obtained from the same DF calculation.}


\rs{In what follows, we demonstrate that our BRSTM simulation approach accurately reproduces the experimentally observed contrast in two paradigmatic systems. 
Firstly, we consider the widely studied molecular semiconductor PTCDA (3,4,9,10-perylene-tetracarboxylic-acid-dianhydride).} 
PTCDA forms flat monolayers with a periodic herringbone structure over metal substrates~\cite{TautzPTCDArev2007,Umbach1996,Tautz2002,RohlfingPTCDASTM2007}, and 
demonstrates various striking effects upon the controlled manipulation with STM on Ag(111)~\cite{Temirov2008Nanotech, Temirov2015PRL,Temirov2018Nature,Temirov2024NatNano}.
Many of those effects arise because the interaction of PTCDA with Ag(111) leads to a partial occupancy of the former LUMO orbital~\cite{Zou2006,Duhm2008,TautzPTCDArev2007,RohlfingPTCDASTM2007} which adds complexity to experimental interpretation and is a significant challenge for simulations.
A combination of the perfectly ordered structure with a wealth of interesting electronic effects resulted in the widespread studies of PTCDA/Ag(111)~\cite{Komolov2000,Wagner2003,Hauschild2005,Zou2006, Komolov2007,Duhm2008,Temirov2006,Schwalb2008,Hauschild2010,Sabitova2018}, and made it into a bench for testing newly developed surface-science techniques, like BRSTM.
Basic features of the molecular structure of PTCDA could be discerned  in the  experimental BRSTM images obtained upon adsorption of PTCDA on Au~\cite{KichinPRB2013} and Ag~\cite{RohlfingPTCDASTM2007, HapalaElectricField2016}, particularly in the central part of the perylene core. However, the observed contrast is mainly related to the chemical composition (e.g. the two C atoms bonded to the oxygens on the two sides of the molecule are absent in the image) and requires a theoretical interpretation~\cite{Krejci2017}.
\rs{Our new series of experimental BRSTM images for the PTCDA/Ag(111) system, taken with just 10~pm of height variation from each other, show a very complex contrast that evolves significantly with the tip height.  Our approach predicts and fully explains the contrast features  and its height evolution assuming that both the $s$ and $p$ channels contribute equally to tunneling. Working with the complete molecule+substrate systems, it accurately captures the effect on the BRSTM contrast of the significant changes in the occupancy of the molecular orbitals upon adsorption, while also highlighting the role of the  CO deflection.}

\rs{Secondly, we address the TOAT/Cu(111) system, where previous simulations with the PP-STM model~\cite{Krejci2017}  failed to reproduce the experimental BRSTM images~\cite{HeijdenACSNano2016}.  A reduction by a factor of 4 of the contribution of the oxygen atoms to the tunneling current, resulting from halving the corresponding tunneling matrix elements,  leaded to a certain improvement, but  the predicted image was still far from the experiment.  Our approach shows the key role of the interaction between adsorbate and substrate in AFM and STM images and the need to include the contribution to the tunneling current of both the  5$\sigma$  and the 2$\pi^*$ CO orbitals in order to reproduce the experimental BRSTM contrast.} 

\section{Results and Discussion}

\subsection{Experimental results}

\begin{figure}[t]
	\centering
	\includegraphics[width=1.0\textwidth]{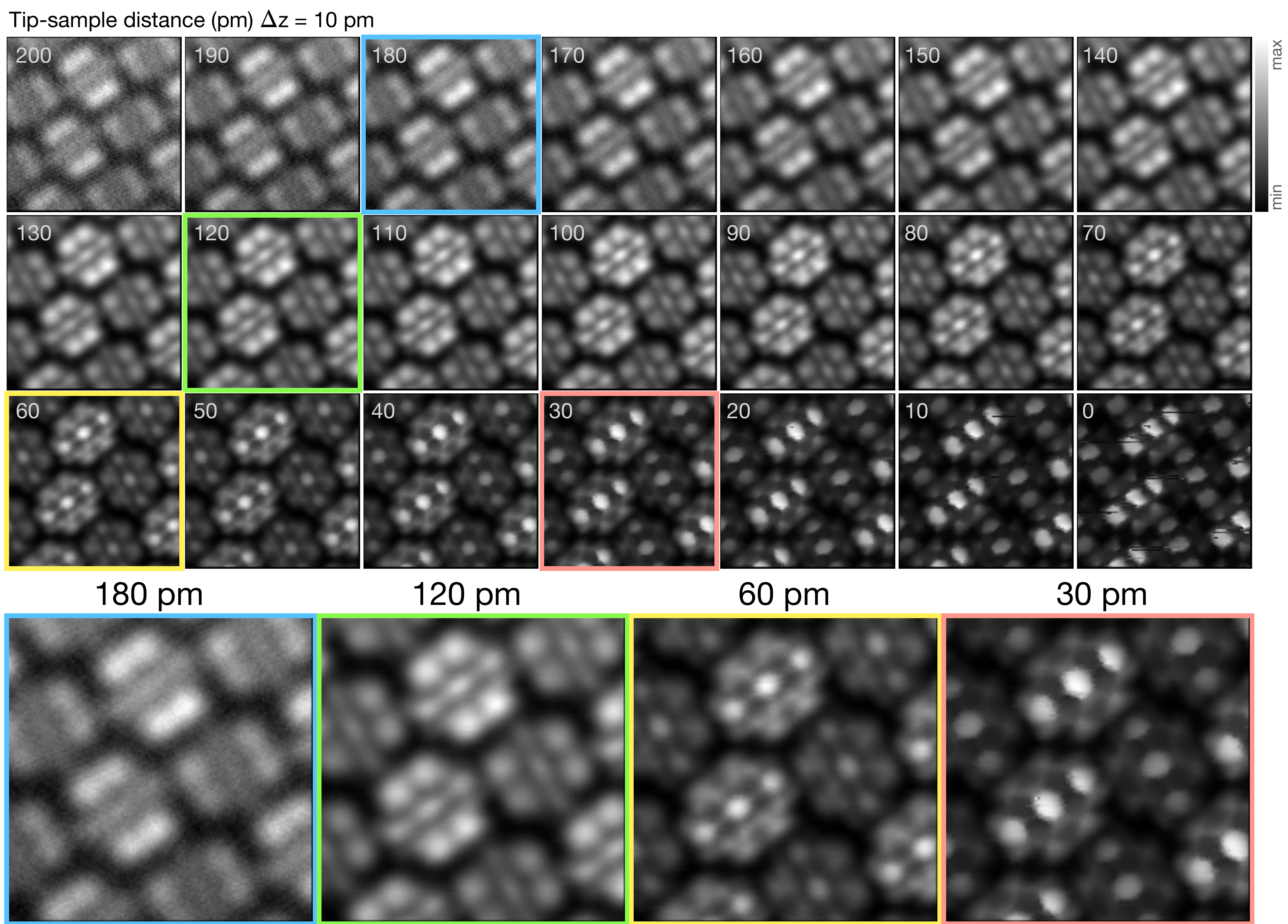}
        \caption{Constant-height dI/dV images (3 nm$\times$3 nm) of PTCDA/Ag(111) measured with a CO-functionalized tip at a bias voltage of -200~mV. The closest tip-sample distance, 0~pm, corresponds to the tunneling set point of $I = 20$~nA and $V = -200$~mV at the center of a bright PTCDA molecule.}
        \label{fig:expstm}
\end{figure}

First, we will review the general characteristics of the BRSTM images of PTCDA/Ag(111) while using a CO-tip.
The images in Figure~\ref{fig:expstm} show differential conductance (dI/dV) in constant-height mode 
recorded with a bias voltage of -200~mV, and thus, they reflect  approximately the shape of the former LUMO of PTCDA. 
These images show the PTCDA herringbone monolayer with two groups of differently  aligned molecules~\cite{RohlfingPTCDASTM2007}.
The two groups can also be identified by a slight contrast difference  (\textit{A} for the brighter and \textit{B} for the darker). 
This difference comes from the electronic interaction of PTCDA with the substrate arising from two adsorption configurations.
However, the most significant information from these images comes from the relaxation of the CO probe, which results in bond resolution.

The effect of the tip relaxation is revealed by the series of consecutive images in Figure~\ref{fig:expstm}, where the tip moves 10~pm closer between images. 
The closest tip-sample distance, 0~pm, corresponds to the tunneling set point of $I = 20$~nA and $V = -200$~mV at the center of a bright PTCDA molecule.

In the images at the furthest distances (e.g., 180~pm, blue border), three lines appear in each molecule: two wide ones along the edges of the molecules and a slimmer one through the middle, with contrast mainly accumulated on the outside lines.
As we go closer and the contrast increases (120~pm, green border), the lines on each side of the molecule separate into two distinct features, while the middle line splits into three different lobes.
At 100~pm, the three central lobes and four lateral features  are almost equally bright. For closer distances,  the three central lobes start dominating the contrast, while the two features on each side form two contiguous triangles pointing away from the molecule, as clearly seen in the image at 60~pm (yellow border).
This reshaping suggests that CO deflection plays a role from this distance onwards.
At 30~pm (red border), clear edges begin to enclose the central lobes in a hexagonal shape due to the strong CO deflection over the bonds in the perylene core of PTCDA.
While this deflection can be explained locally by the covalent bonds within the molecules, the substrate can affect the image's overall appearance in this system.

\subsection{Effect of the substrate in PTCDA/Ag(111)}\label{sec:subeffect}

\begin{figure}[t]
        \centering
        \includegraphics[width=0.6\textwidth]{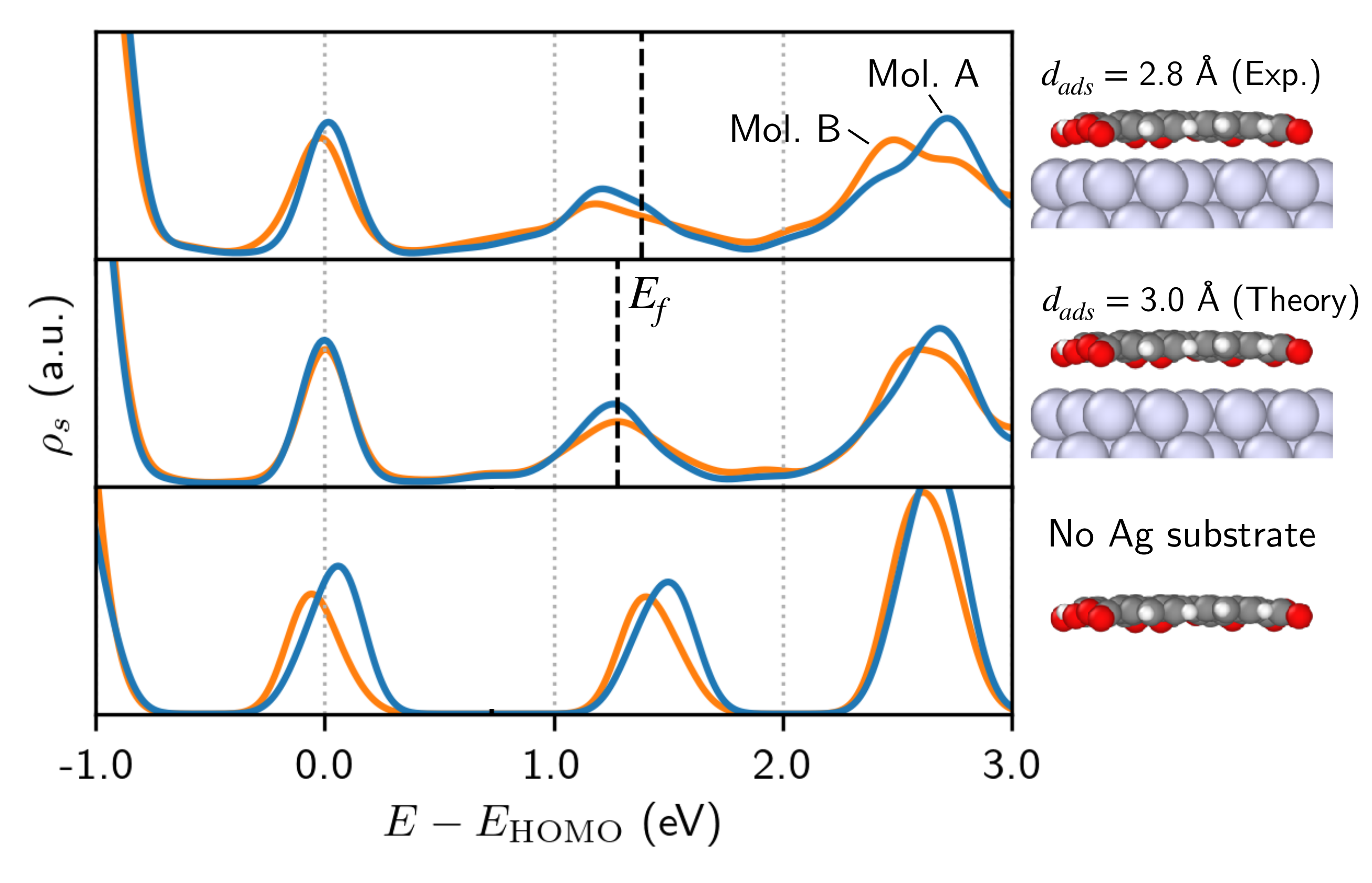}
        \caption{Theoretical projected density of states of the PTCDA monolayer on Ag(111) (\textit{A} molecule in blue and \textit{B} in orange) with respect to the HOMO level (\(E_{\mathrm{HOMO}}\)), the Fermi level is represented by the vertical discontinuous lines. The top panel shows the pDOS for a DFT calculation using the experimental adsorption distance, the middle with the distance predicted using the DFT-D3 approximation for the vdW interaction, and the bottom one corresponds to the PTCDA monolayer in their relaxed adsorption configuration but without the metal substrate. Only the calculations at the experimental adsorption distance reproduce the position of the partially occupied LUMO found in the experiments.}
        \label{fig:pdos}
\end{figure}

The contrast difference between the \textit{A} and \textit{B} molecules in the experimental images is evidence of the effect of the Ag substrate on the electronic structure of PTCDA.
In STM measurements, this is associated with a different partial filling of PTCDA's LUMO~\cite{RohlfingPTCDASTM2007}.
To identify this effect in the electronic structure, we extract from the  DFT calculations (see Methods for technical details) the Projected Density-of-States (pDOS) on each molecule.
The bottom panel of Figure~\ref{fig:pdos} shows the pDOS of the PTCDA monolayer calculated for its relaxed adsorption configuration but not including the metal substrate in the calculation. As expected, the HOMO states of both molecules are occupied, and the Fermi level  (\(E_f\)) is located in between the HOMO and the empty LUMO states.
In contrast, the middle and top panels show the monolayer with the substrate (all referenced to the molecular HOMO).
Here, we can observe a significant change in the system's Fermi level, as, in these two cases, the LUMO states of PTCDA are partly filled, as measured in the STM experiments.

However, our DFT calculations --combining the PBE exchange and correlation (XC) functional and the semi-empirical DFT-D3 correction for the dispersion interaction-- reproduce neither the position of the LUMO states ($\sim$0.2-0.3 eV below $E_f$ according to the STM experiments)~\cite{RohlfingPTCDASTM2007} nor the difference between A and B molecules. 
These shortcomings come from a  known limitation of modern approaches that combine gradient-corrected XC functionals, like PBE, with an approximation for the long-range dispersive interactions, in our case DFT-D3, to describe the molecule-metal substrate interactions. 
Although widely used for simulating molecular adsorption, our PBE+DFT-D3 methodology and other more sophisticated approaches tend to underestimate the molecule-substrate interaction and, thus, overestimate the adsorption distance (\(d_{\mathrm{ads}}\)) (see ref.~\onlinecite{Hormann2020JCP} for a thorough discussion of the PTCDA/Ag case). The interaction is rather flat in the relevant distance range and even small errors in the energy can lead to large differences in the predicted adsorption distance. 
In the PTCDA/Ag(111)  case, the \(d_{\mathrm{ads}}\) ($\sim$ 300~pm, defined as the average height of all atoms in both PTCDA molecules with respect to the metal surface plane) calculated with our  PBE+DFT-D3 combination  is larger than the experimental one ($\sim$ 280~pm). This experimental distance has been accurately characterized 
with x-ray standing waves~\cite{Hauschild2005, Hauschild2010}. 
As none of the theoretical approaches provide a  \(d_{\mathrm{ads}}\) close to the experimental value~\cite{Hormann2020JCP}, 
we manually placed the PTCDA molecules closer to the surface so that their adsorption distance corresponded to the experimental value of  \(d_{\mathrm{ads}}\)  and then used the PBE+DFT-D3 methodology to calculate the electronic structure of this system.
Then, we have relaxed the whole system to its ground-state configuration  keeping fixed the $z$-position of the C atoms in the perylene core of each PTCDA molecule.

The resulting pDOS, in the top panel of Figure\ref{fig:pdos}, shows that the LUMOs move now to a position closer to the experiments and the difference in the corresponding pDOS for the two molecules becomes larger than in the case where the PBE+DFT-D3 \(d_{\mathrm{ads}}\) was used (mid panel).
While this last change may seem small, it greatly improves the theoretical description of the contrast difference between the \textit{A} and \textit{B} molecules. However, the agreement is not perfect as, according to the experimental STS spectra~\cite{RohlfingPTCDASTM2007},  the pDOS of the two molecules should be shifted with respect to each other,  with the maximum for the A (B) molecule around 0.2 (0.3 eV) below  $E_f$.  

 \begin{figure}[t]
	\centering
	\includegraphics[width=1.0\textwidth]{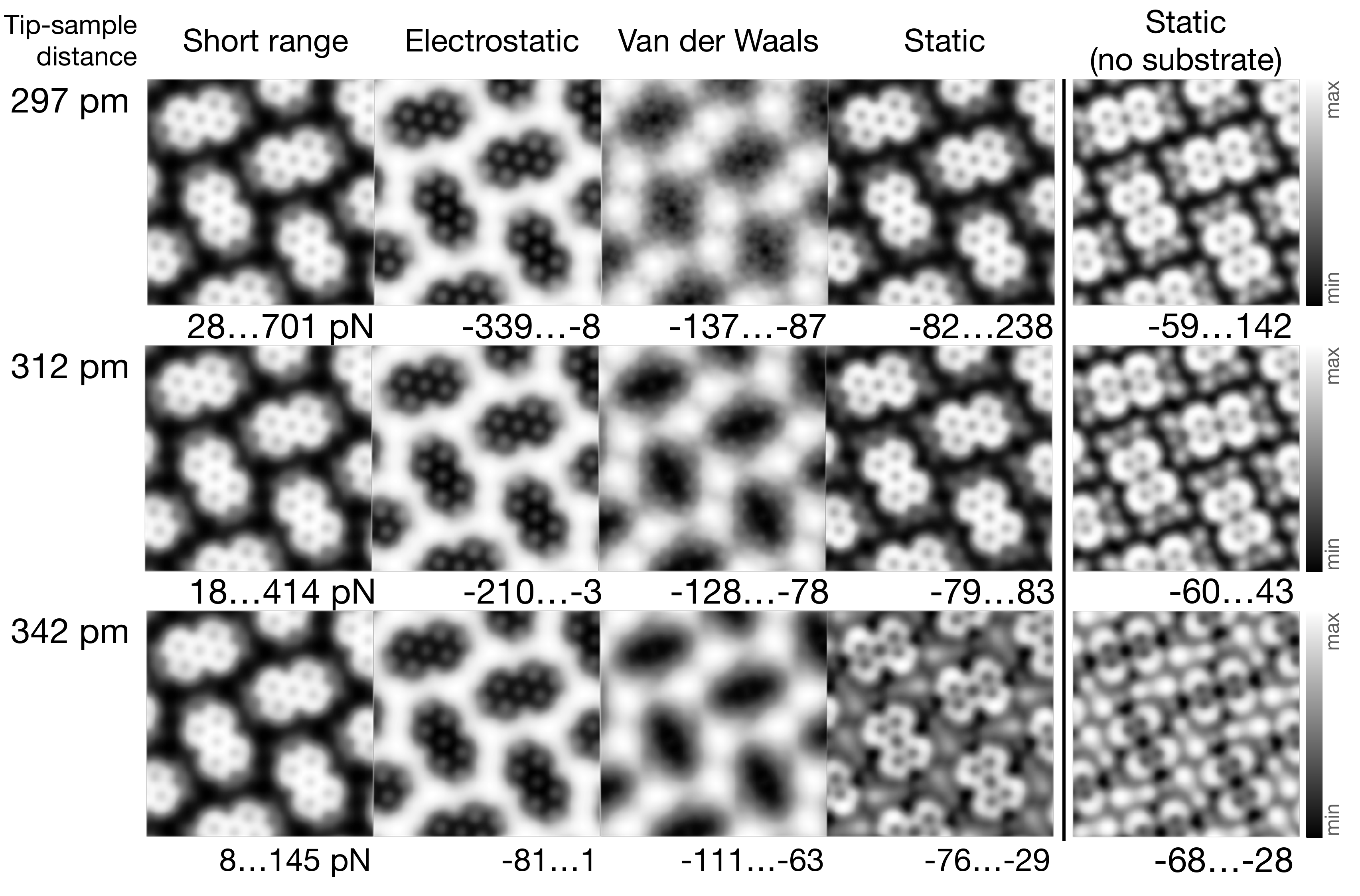}
        \caption{Individual force contributions (short-range Pauli repulsion, electrostatics and vdW --see Figure~\ref{fig:framework})  and total sum (labelled Static) to the theoretical AFM image of FDBM at different tip-sample heights for PTCDA/Ag(111), taking as a reference the average position of the C atoms in the perylene core of both PTCDA molecules. The last column shows the total static image of the PTCDA monolayer without substrate. The gray scale is adjusted to the maximum and minimum value of each image, shown in the labels below each one (positive is repulsive and negative attractive).}
        \label{fig:interactions}
\end{figure}

The change in electronic structure induced by the interaction with the substrate  means that the charge density will now include a more significant contribution from the LUMO. In turn, the total repulsion and the CO deflection will be significantly different from the one obtained just including the PTCDA molecules in the DFT calculation. 
Thus, including the substrate is crucial to get an appropriate simulation of the CO deflection over our system, which is one of the two critical components of BRSTM simulations.
FDBM can capture these changes correctly when calculating the complete charge density of the substrate/sample system \textit{ab-initio} (DFT). To illustrate this, we compare the static tip-sample FDBM force \(F_{\rm{static}}\) --determined as a numerical derivative of the static  (i.e., non-deflected probe) tip-sample potential \(V_{\rm{static}}\), see Methods-- calculated 
with or without the substrate.

The static force aggregates the interactions in the first three columns of Figure~\ref{fig:interactions}: the Pauli repulsion, electrostatic  and Van der Waals interactions.
For PTCDA/Ag(111), the static force maps in the second-to-last column of Figure~\ref{fig:interactions} show that contrast accumulates in the perylene core.
This feature is maintained in the static image calculated  without substrate (last column); however, there is a clear difference in the oxygens on the ends of PTCDA.
When including the substrate, they lack contrast because of their chemical interaction with the atoms in the silver surface.
On the other hand, when not including the substrate, the three oxygens on each end of PTCDA can be more easily identified because their strongly localized charge. 
Furthermore, at the largest tip-sample separation, there are also changes to the perylene core contrast.
This occurs because charge accumulates slightly more on the long hydrogen-terminated sides of the perylene core, instead of being homogeneous over the whole core.
Lastly, the maximum repulsive force for each image plane is considerably larger when including the substrate.
While this may not change the qualitative appearance of the simulated HRAFM images, it is very important to have an accurate description of the probe deflection for consecutive tip-sample separation.
As such, we have included the substrate in all of the HRAFM simulations.


\subsection{Relaxation of the CO-probe}

After obtaining \(F_{static}\) and understanding the role of the substrate in the tip-sample interaction, we can calculate the CO deflection at every point of the HRAFM image.
To do this, we consider that the tip-C-O system is effectively a torsion-spring lever where the O is limited to move in spherical coordinates (\(\theta\) and \(\varphi\)).
The restoring force of the spring acts on \(\theta\) (see Figure~\ref{fig:framework} and Methods).

\begin{figure}[t]
	\centering
	\includegraphics[width=0.7\textwidth]{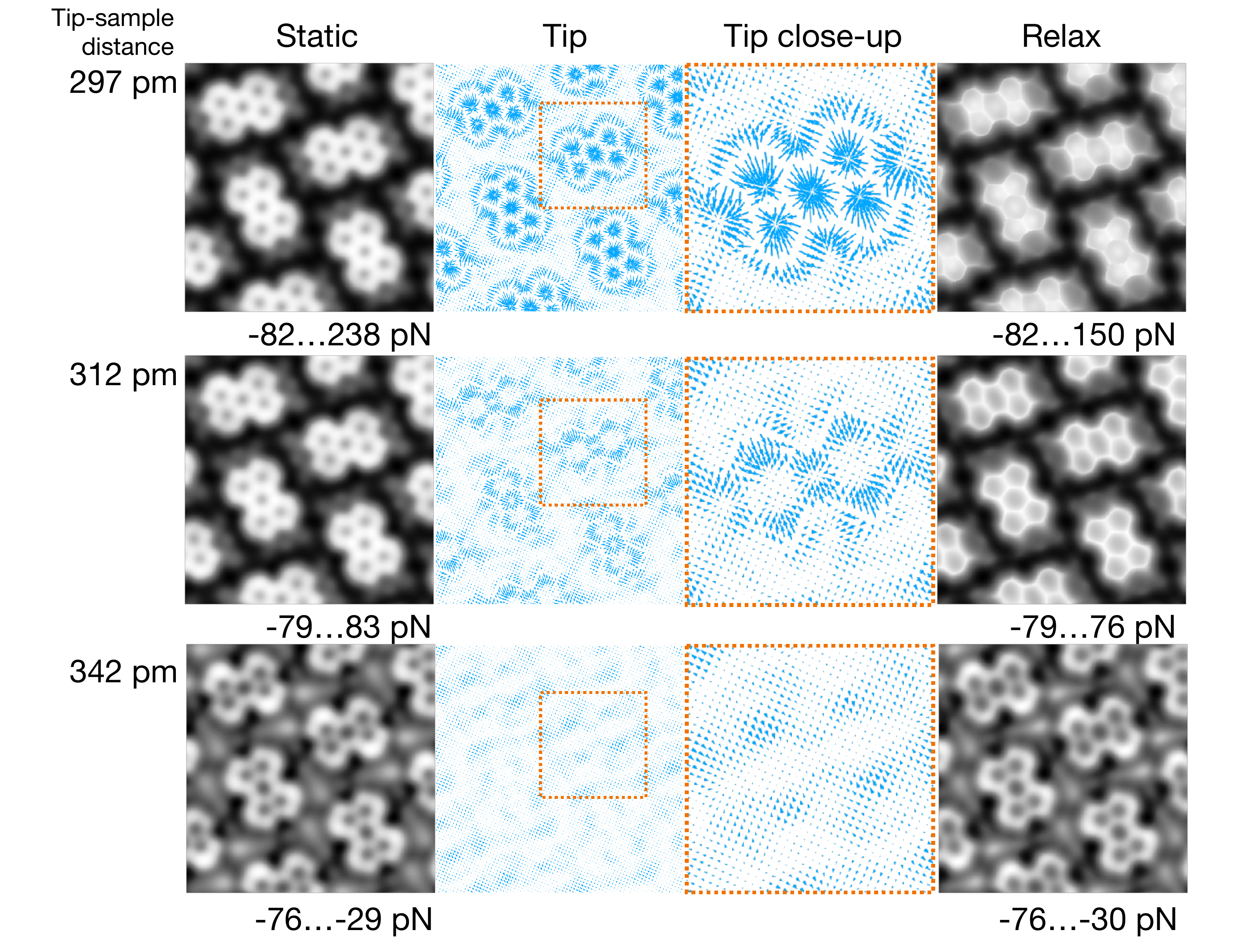}
        \caption{HRAFM images of PTCDA/Ag(111) with the static force map (left), the deflection of the CO on the tip on the static force map, shown as the shifts (2:3 scale) of the O atom  projected onto the \textit{xy}-plane  (middle left), close-up of the area  delimited by the orange rectangle (middle right), \textit{relaxed} HRAFM images accounting for the CO deflections (right). The gray scale is adjusted to the maximum and minimum value of each image, shown in the labels below each one (positive corresponds to a repulsive force and negative to an attractive one).         }
        \label{fig:tippos}
\end{figure}

The middle column of Figure~\ref{fig:tippos} shows the relaxed position of the O in the CO-tip projected onto the \textit{xy}-plane.
At the closest tip-sample distance, 297~pm, 
we can see that much of the tip deflection is pointing into the centers of the carbon rings of the perylene core and, to a lesser extent, toward the outside of the PTCDA molecules.
These features help the interpretation of the \textit{relaxed} HRAFM images in the right column of Figure~\ref{fig:tippos}.
%
There, we can see that the image of the perylene core sharpens, and a faint structure on both ends of the molecules suggests the presence of the terminal oxygen atoms.
%
The contrast in the center of the carbon rings, characteristic of HRAFM images at very close distances,  reflects that  although the maximum repulsive force is significantly reduced (by ~37\%) in the relaxed image, it is still very high and tend to saturate the gray scale.  

There is a proportional decrease in tip relaxation when moving further away from the sample.
For the 312~pm separation, the maximum force is reduced from 83~pN in the static image to 76~pN in the relaxed one.
Nevertheless, the internal structure of PTCDA is still very well defined, with the contrast inside the rings disappearing due to the effect of the smaller difference between the  minimum and maximum forces (-79~pN  and 76~pN,  compared to -82~pN  and 150~pN for the 297~pm separation) in the gray scale.
After 342~pm of separation, there is barely any change in the relaxed images as the corrugation of the  potential-energy surface in \(V_{\rm{static}}\) is insufficient to deflect the CO probe.
However, there is a very slight tilting of the CO probe towards the more attractive regions in the center of the carbon rings, explaining the larger minimum attractive force (-30~pN) in the relaxed image compared to the static one (-29~pN). 

\subsection{Simulation of \textit{s} and \textit{p}-wave STM channels of PTCDA/Ag(111)}

\begin{figure}[t]
	\centering
	\includegraphics[width=1.0\textwidth]{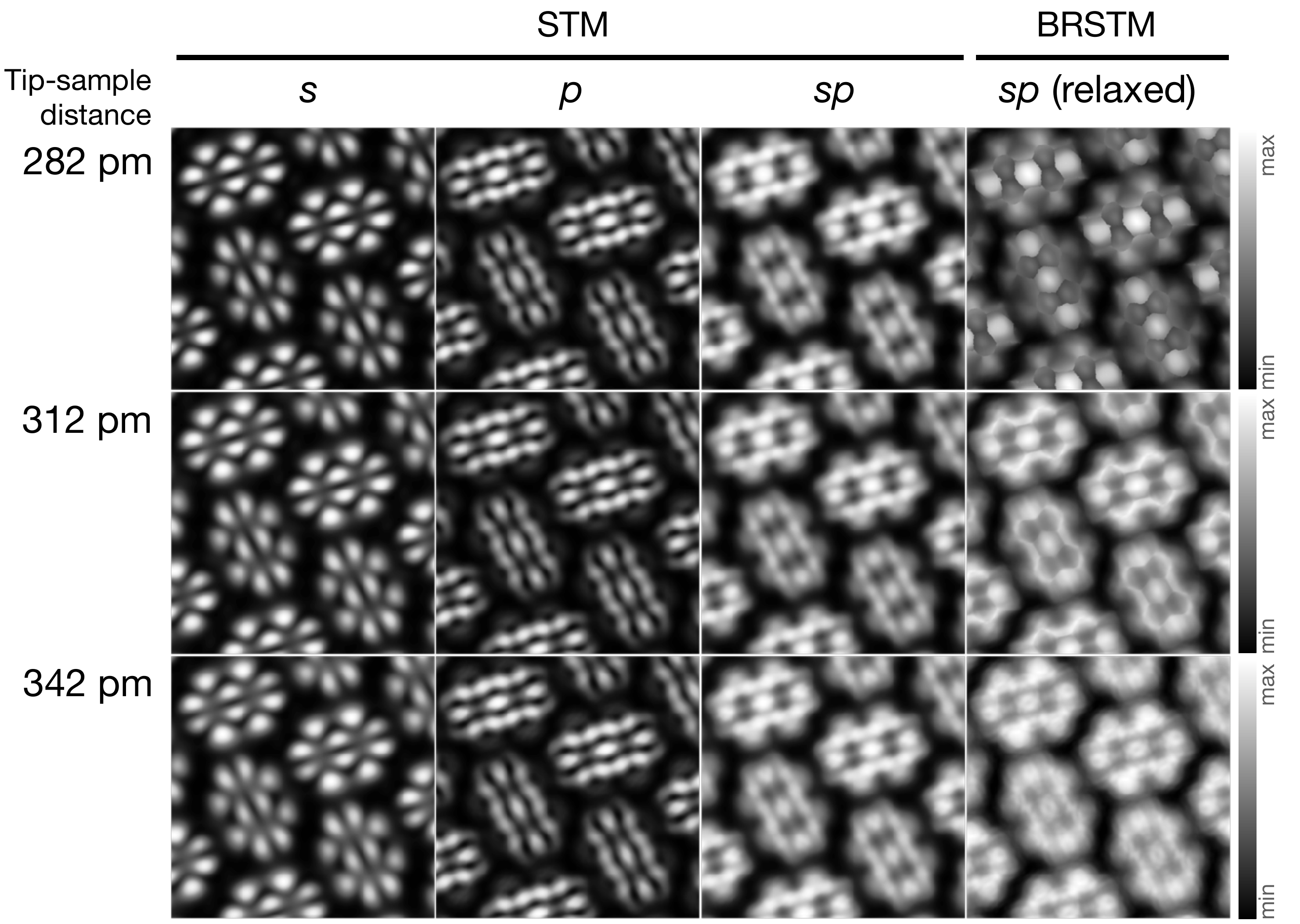}
        \caption{Simulated STM images with a CO-tip of PTCDA/Ag(111) with a bias of -0.25~V and different tip-sample distances. They include the \textit{s}-wave channel (left), the \textit{p}-wave (middle left), the combined \textit{sp} images with a 1:1 ratio (middle right), and the relaxed BRSTM images (right). The gray scale is adjusted to the maximum and minimum value of each image.}
        \label{fig:thstm}
\end{figure}

Now that we have obtained the CO deflections 
--the first ingredient of BRSTM images with a CO-tip--, we will delve into obtaining the tunneling signal with the CO orbitals.
%
To this end, we used the DFT wave functions we had already calculated to obtain the local density of states and the sample charge density in a 3D grid, integrating the states between $E_f$ and -0.25 eV ( the position of the maximum of the pDOS for the A molecule close to $E_f$).. 
Following the work of Gross et al.~\cite{gross2011b}, we account for the \textit{s}-wave component (representing the tunneling from the CO $5\sigma$ orbital)  using  the Tersoff--Hamann approach, $I_s \sim |\psi_{\rm{sample}}|^2$, while the \textit{p}-wave component  (accounting for the tunneling from the $2\pi^{\ast}$ orbital) with the Chen equations for the \(p_x\) and \(p_y\) orbitals, $ I_p \sim  \left|\frac{\partial \psi_{\rm{sample}}(\mathbf{r})}{\partial x} \right|^2 +
    \left|\frac{\partial \psi_{\rm{sample}}(\mathbf{r})}{\partial y} \right|^2 $ (see Methods). 

The first two columns of Figure~\ref{fig:thstm} show the \textit{s} and \textit{p} channels of the simulated STM images with a -0.25~V bias at three different tip-sample distances.
The \textit{s}-channel represents almost exactly the PTCDA LUMO, 
reflecting the partial occupation of this orbital upon adsorption on Ag(111) discussed above.  
Also, we can see the contrast difference between the A and B molecules.
For the \textit{p}-wave channel, the location of the features corresponds to the nodes between the lobes of the PTCDA LUMO.
These are three large lobes along the long axis of PTCDA, coming from the three central pairs of lobes in the LUMO, and four small lobes on each side, stemming from the nodes between the two external LUMO lobes and the three central ones on each side.

The third column of Figure~\ref{fig:thstm} shows the combined \textit{s} and \textit{p} channels with a 1:1 sum.
We chose these relative weights because Gross et al.~\cite{gross2011b} identified them as a suitable combination to reproduce their experimental data. Our results also exhibit strong concordance with this combination (see Figure~\ref{fig:th-exp} and the discussion below).
The images show the central pairs of lobes of the \textit{s}-channel and the single ones of the \textit{p} combining to form the three conspicuous features in the center of the molecule.
At the same time, the external lobes of the \textit{s}-channel combine with the small lobes on the sides of the \textit{p} signal to form a continuous feature on each side of PTCDA with two triangular shapes pointing away from the molecule.
%
A recent STM/AFM study on pentacene and  naphthalocyanine on bilayer NaCl on Cu(111) revealed a transition from predominant  \textit{p} to  \textit{s}-wave tip contrast upon increasing the tip-sample distance~\cite{Paschke2025}. The distance-dependent contrast change was explained by the steeper decay of the tunneling matrix element for tunneling between two p-wave centers, compared to tunneling between two s-wave centers. Using  a tip with a fixed \textit{s}:\textit{p} ratio of 2.5:1, simulations could reproduce the experimental data, including the distance-dependent transition from predominant \textit{p} to  \textit{s}-wave tunneling contribution. While this ratio seems to be  suitable for the large tip heights considered in that study, in the range of 360 to 900 pm,   our work focuses on a much closer range, between 280 and 350~pm, where the effects of CO relaxation are important. In this range, the p-wave contribution is expected to have a very important role, explaining the success of a 1:1 mixing in reproducing our experimental results for PTCDA.

Now, we use a tri-cubic interpolation of the 3D grid of the simulated \textit{sp} STM images over the CO positions obtained with FDBM.
The right-hand side column of Figure~\ref{fig:thstm} shows the interpolation results: a \textit{relaxed} BRSTM image with a CO-tip.
The relaxed BRSTM images show the same tip-sample distance dependence of the experimental images and the HRAFM simulation.
For the furthest distance, 342~pm, the only noticeable change, because of the minimal CO relaxation, is the slightly more homogeneous structure of the four triangular features on the sides of PTCDA.
At 312~pm, the CO deflection plays a more significant role, as the central lobes are enclosed in a defined hexagonal shape, and the triangular features are shaped as arrows pointing away from the sides.
At the closest distance, 282~pm, the central lobes dominate the contrast and are entirely enclosed in the hexagons of the PTCDA rings.
Also, other rings in the molecule can be distinguished because the CO tip is deflected away from bonds, where there are slight jumps in contrast as the CO suddenly changes location (hence the name bond-resolved STM).

\subsection{Theoretical and experimental comparison}

\begin{figure}[t]
	\centering
	\includegraphics[width=1.0\textwidth]{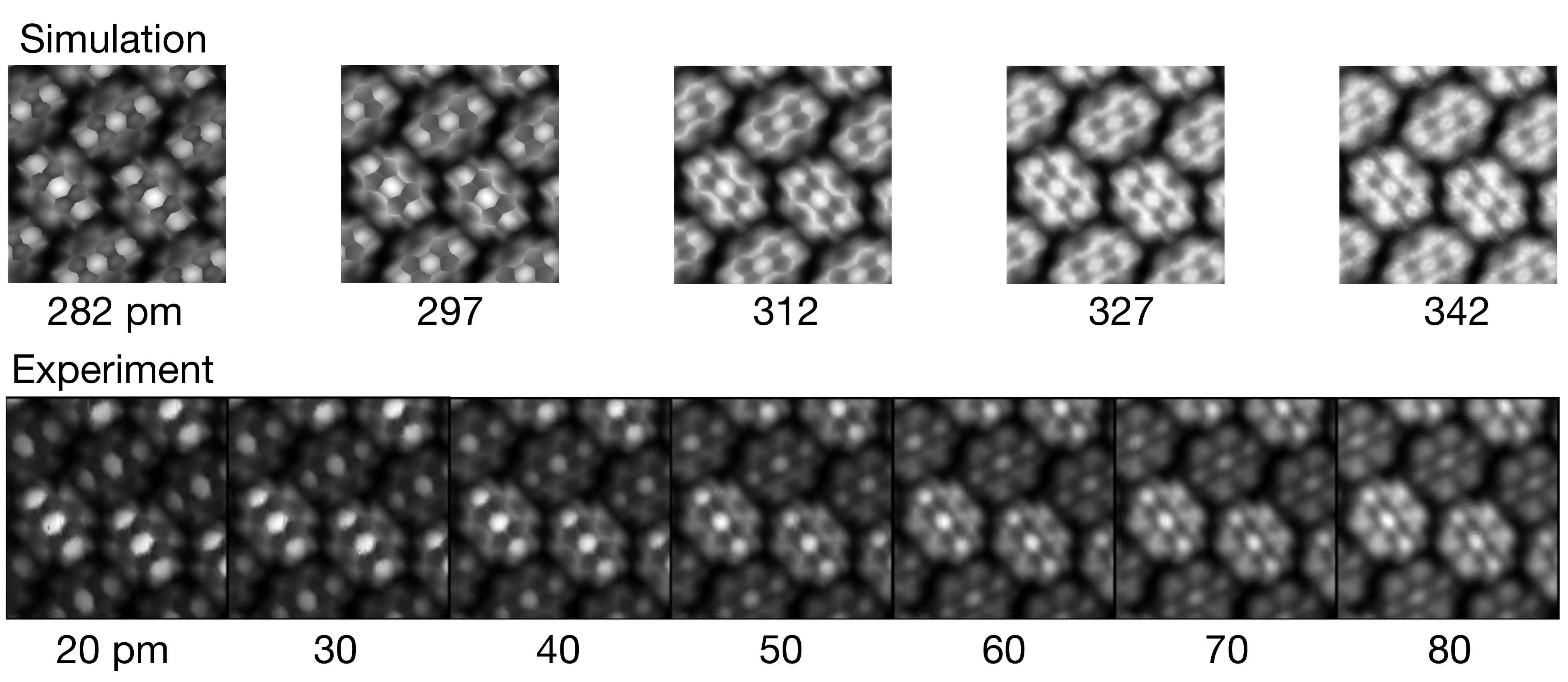}
        \caption{Theoretical (top) and experimental (bottom) BRSTM images of PTCDA/Ag(111). Details as in Figures~\ref{fig:thstm} and \ref{fig:expstm}, respectively.}
        \label{fig:th-exp}
\end{figure}

We can now compare a sequence of theoretical STM images to the experimental sequence to corroborate the accuracy of the calculations.
We selected the images in Figure~\ref{fig:th-exp} to match the tip-sample distances between theory and experiments, concluding that the experimental reference corresponds to $\sim 262$~pm.
The general structure of experimental and theoretical images is very similar.
There are three bright central lobes in a single line and two pairs of triangular features on either side of the molecules.
For the closest theoretical image, with a 282~pm distance, the best experimental image is the 20~pm one. Thus, we conclude that the experimental reference corresponds to $\sim 262$~pm. With this match, there is a very good agreement between theory and experiment in the contrast evolution with the distance. 

%
In the images at 282~pm, we see the central lobes enclosed in the hexagonal-shaped rings of PTCDA.
Also, the contrast in the middle lobe appears homogeneous, while in the side lobes, it fades toward the ends of the molecule.
Furthermore, the central lobe appears enclosed by four 
other rings with much less contrast that have an irregular hexagonal shape.
Lastly, the contrast in the sides of the molecules appears to spread out from two points on each side.
In the middle distance, 312 and 50~pm for theory and experiment respectively, the central lobes appear more diffused, but the ones on the ends still seem to fade towards the outside.
Additionally, the two features on each side appear again as two arrows defined by a single line on each side.
Finally, for the furthest distance, 342~pm for theory and 80~pm in the experiment, all features are less defined as there is minimal CO-tip deflection.
Still, the theoretical and experimental images share the central lobes' shape and the side features' triangular shape. 
\rp{Note that, while there is a great match in the contrast evolution, the alternating brightness of molecules observed experimentally at all the distances is not fully captured in the simulations, probably  due to the limitations of the DFT description of the adsorption geometry, discussed above, which affects the local density of states of the two molecular types.}

\subsection{A challenging system: TOAT/Cu(111)} 

After showing the ability of our model to describe the complex BRSTM contrast and its strong distance dependence in the case of PTCDA/Ag(111), we focused our efforts on understanding a challenging system, TOAT adsorbed on Cu(111), where experimental data was available in the literature~\cite{HeijdenACSNano2016} and previous theoretical attempts to explain the observed contrast failed~\cite{Krejci2017}.

\begin{figure}[t]
        \centering
        \includegraphics[width=0.8\textwidth]{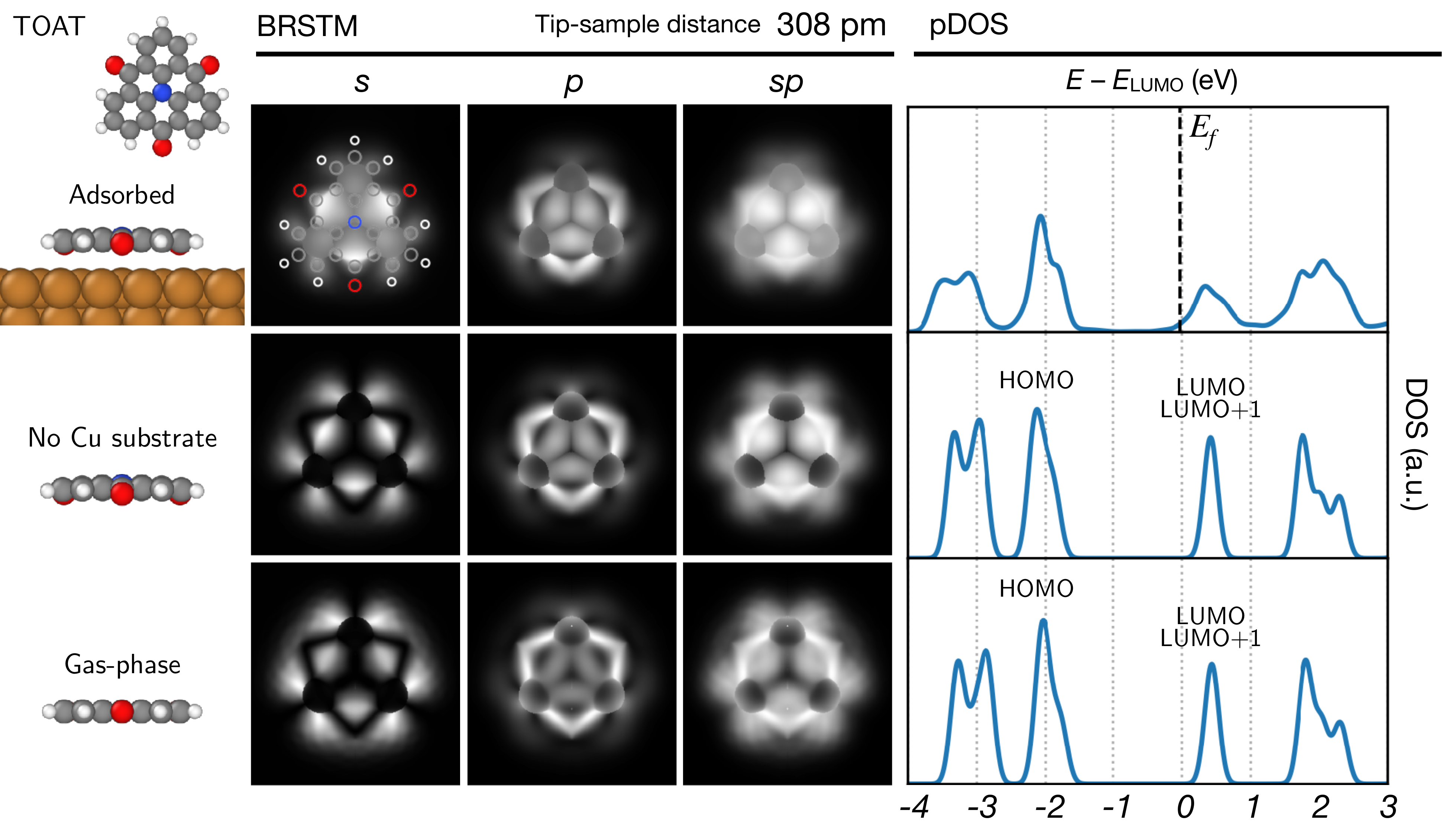}
        \caption{Simulated BRSTM images and projected Density of States (pDOS) of different TOAT in different configurations. Their atomic geometry (in side view) is shown on the left. The BRSTM images (left panel) include the LUMO and LUMO+1 levels and where calculated  at 308~pm of tip-sample separation. From top to bottom, the simulated systems are the adsorbed TOAT over Cu(111), the TOAT in its adsorption configuration but without the metal substrate, and the gas phase TOAT. From left to right, the columns show the BRSTM \textit{s}-wave channel, the \textit{p}-wave, and the combined \textit{sp} channels. The gray scale is adjusted to the maximum and minimum value of each image. Theoretical pDOS of a TOAT molecule with respect to the LUMO level (\(E_{\mathrm{LUMO}}\)). The top panel shows the pDOS of a calculation with both molecule and substrate, with the Fermi level --represented by the vertical discontinuous line-- located right before the onset of the LUMO and LUMO+1 states, that remain empty in the adsorbed system.} 
        \label{fig:toatdos}
\end{figure}

Figure~\ref{fig:toatdos} compares the results of three different DFT calculations (in particular, the pDOS) and their associated  BRSTM simulations (separating the contribution of the different tunneling channels). The corresponding probe-deflection maps and HRAFM images can be found in the Supporting Information (Figure~S1).
Our reference (panels of the top row in Figure~\ref{fig:toatdos}) is the TOAT molecule adsorbed over the Cu(111) substrate. While planar in the gas phase, upon adsorption,  the molecule is slightly deformed, 
with the atoms in the outside edges bending slightly downwards due to the interaction with the metal substrate (top panel, first column of Figure~\ref{fig:toatdos}). The three O atoms (located right on top of Cu atoms) are the closest to the surface, followed by their neighboring C's, with the C atoms on the opposite vertex only slightly above. In order to disentangle the effect in the BRSTM images of the substrate-induced changes in the electronic properties and in the geometry, we consider two additional cases:  the molecule in its adsorbed (deformed) configuration but without the substrate (panels in the middle row in Figure~\ref{fig:toatdos}), and the TOAT molecule in its gas-phase, planar structure (bottom row).

According to the pDOS for the TOAT/Cu(111) case (top panel in the last column of  Figure~\ref{fig:toatdos}), its Fermi level is located precisely at the onset of the unoccupied molecular states. Comparing with the pDOS for the two other cases, we see that DOS associated with the HOMO, LUMO and LUMO+1 states (note that LUMO and LUMO+1 are degenerate in TOAT) has broadened due to the electronic overlap with the metal wavefunctions, but the location of these molecular orbitals has not changed significantly compared to the isolated molecule. At variance with the PTCDA/Ag(111) case, the LUMO and LUMO+1 remain empty, preserving the occupancy of the molecular states. This makes  TOAT/Cu(111) a very good candidate  for assessing the effect of the substrate on the BRSTM images in the cases where the molecule-metal interaction is weak and does not have a significant effect in the molecular electronic structure. 

BRSTM images for a tip-sample distance of 308~pm (where the CO deflection is already relevant, see Figure~S1 in the Supporting Information) are displayed in the central three columns of Figure~\ref{fig:toatdos}. Following the experiments~\cite{HeijdenACSNano2016}, they correspond to tunneling into the empty states in the TOAT degenerate LUMO and LUMO+1.
In the case of the adsorbed TOAT (molecule and substrate are included in the calculation), the images include the states between the Fermi level and +200~mV.
%
In the two cases that do not consider the substrate, the BRSTM contrast is obtained integrating over the sharp peak in the pDOS associated with the LUMO and LUMO+1.  
Notice that the structural changes induced by the adsorption have a very minor influence on the pDOS, although they do play a significant role in the probe deflection. 

The BRSTM images of the \textit{s}-channel give us a clear idea of the role of the substrate. These images are essentially proportional to the square of the wavefunctions integrated over the relevant energy range.
Starting with the images 
of the two  molecular structures without the substrate (panels in the middle and bottom rows), 
there are two areas with significant contrast that repeat three times following the three-fold symmetry of the molecule. The first one is defined by the C atom directly bonded to one of the O atoms (the O atom itself is missing from the image), and the two C atoms connected to this carbon in the ring. The combined effect results in a bright triangular feature. The other area is associated with two of the three C atoms, saturated with H atoms, in one of the outer benzene rings. These two atoms, bonded to the C atom in the vertex (missing in the image), give rise to two rather localized features along each of the  C-H bonds. 
All of these features have significant contrast as the probe relaxation brings out very defined edges and the features are far apart from each other (see SI, Figure~S1 where show the probe relaxation and compare the STM and BRSTM images).

\begin{figure}[t]
        \centering
        \includegraphics[width=0.7\textwidth]{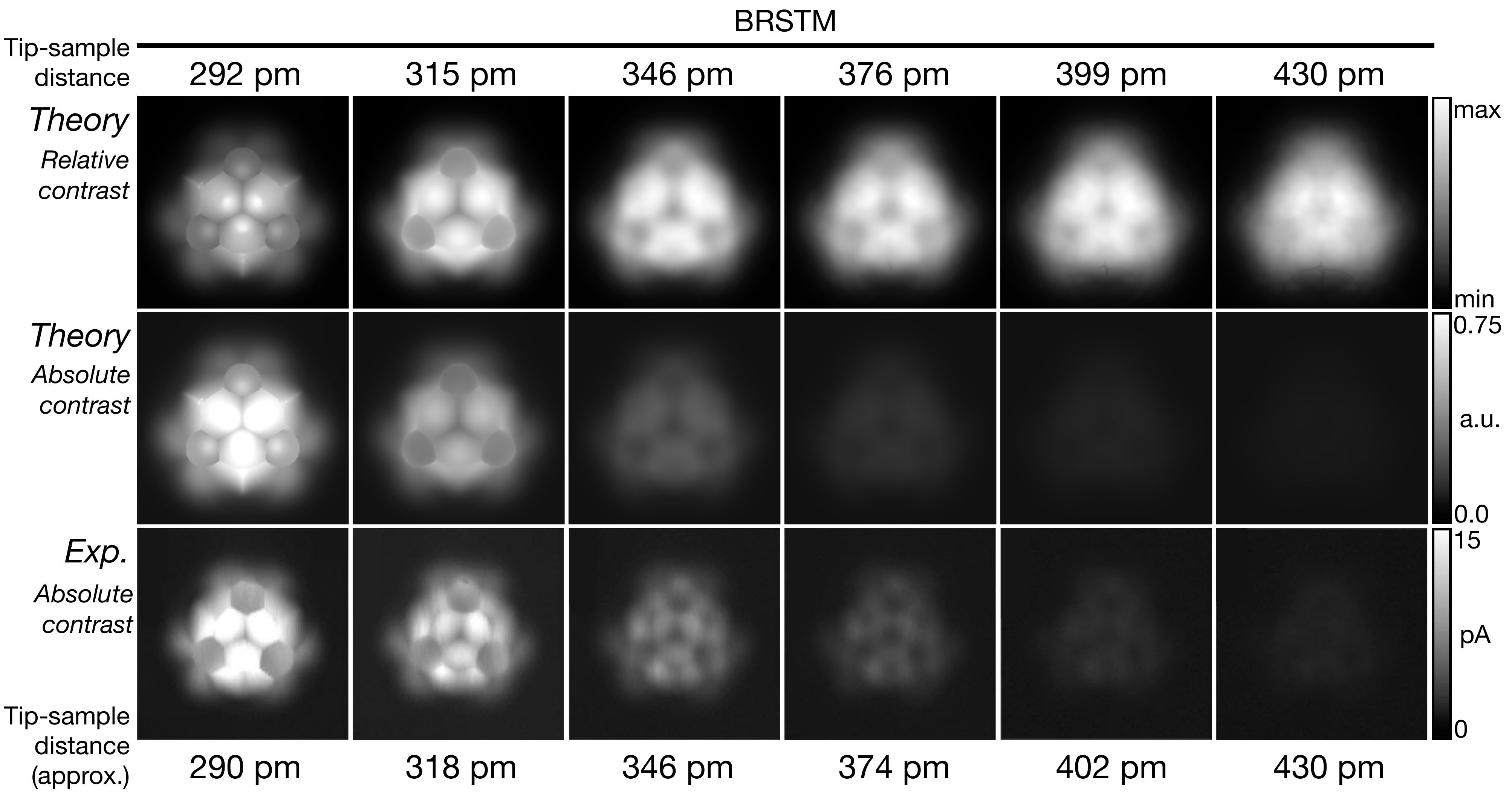}
        \caption{BRSTM images of TOAT/Cu(111).  The top and middle rows show the simulated images and the bottom row the experimental measurements. 
From left to right, the columns show images for an increasing tip-sample distance.  Experimental distances have been referenced by matching the contrast evolution with the theoretical simulations. 
The images in the top row are depicted with relative contrast, that is, the gray scale is normalized and mapped for each tip-sample distance. To allow a direct comparison with the experiments, the theoretical images in the middle row have the same gray scale, where the intensity was normalized across all of the shown tip-sample distances and has a saturation level of 0.75, while the experimental images have a saturation level of 15~pA (more details in ref.~\protect\onlinecite{HeijdenACSNano2016}). 
Experimental images were adapted with permission from van der Heijden, et al., {\textit ACS Nano} 2016, 10, 8517--8525. Copyright 2016 American Chemical Society.}
        \label{fig:toatbrstm}
\end{figure}

However, including the metallic substrate reduces the contrast by adding a faint signal across the whole molecule.
This background signal markedly changes the structure of the images, ``illuminating'' previously dark areas in the center of the molecule and the outer rings and resolving the ring structures in these areas as the probe deflection along the C-C bonds is now much more noticeable.
On the other hand, the \textit{p}-channel has more structure than the \textit{s} in all cases because its contrast is associated with the gradients of the wavefunctions in the \textit{xy}-plane (see Methods). These gradients vary more rapidly in space than the wavefunctions themselves. Also, due to this dependence in the gradients, the brightest features in the $p$-channel  correspond to the dark areas in the  $s$-channel. 
In the final image, when both channels are combined with a 1:1 weight, the adsorbed molecule with the metal substrate  has more contrast and structure than the two free molecule cases, bringing the theoretical simulation to a very good agreement with the observed experimental contrast (see bottom row in Figure~\ref{fig:toatbrstm}).
Comparing the middle and bottom rows in Figure~\ref{fig:toatdos}, it is clear that including the molecular distortion induced by the adsorption yields a better image, with more structure in the inner rings than the flat, gas-phase molecule. However, the inside of the three outer rings still has no signal. It is only when we include the substrate in the calculation that we fix these limitations and obtain a simulated BRSTM image in good agreement with the experiment. 

Figure~\ref{fig:toatbrstm} shows, in the top row,  the contrast evolution in the simulated BRSTM images for the TOAT/Cu(111) system as a function of the tip--sample distance. The distances have been chosen to facilitate the comparison with the experimental measurements~\cite{HeijdenACSNano2016}, displayed in the bottom row. Notice that the distance in  Figure~\ref{fig:toatdos}, 308~pm, lays between the first two distances in this figure, 292 and 315~pm. In the top row, each image used a relative gray scale, where black corresponds to the minimum value of the current, $I$, and white to the maximum value within this particular image. This representation is useful to maximize the contrast in each image, but masks the strong (exponential) distance dependence of the tunneling current. The panels in the middle row correspond to these same images but using a fixed absolute gray scale, that allows a more faithful comparison with the experimental images, where the same gray scale (with a 0-15~pA range) is used in all the images. This is really important when comparing the relative brightness of different features in the same image.
The match between theory and experiment is even better using a fixed gray scale, and allow us to determine an absolute tip-sample distance for the experimental tip position at closest approach. Notice that, with this fixed gray scale, the simulations capture very well the halos that appear around the position of the O atoms and  the marked central feature associated with the C-O bond, and highlighted by the probe deflection. These images further support that the probe deflection only plays a role for the  closest tip-sample distances, below 315~pm, and that the signal from the \textit{s}-channel is crucial to capture the experimental contrast.

\section{Conclusions}

We have developed an ab-initio based approach for simulating BRSTM images and validated its performance by comparing its results to experimental data. The set of experimental BRSTM images of PTCDA/Ag(111)  was collected using a CO-functionalised STM tip changing the tip-surface distance with 10 pm steps, thus yielding the BRSTM contrast that spans a wider range of tip-sample distances. 
Our simulated BRSTM images closely match our experimental results as well as the BRSTM data for TOAT/Cu(111)  discussed earlier in the literature. 
Our approach combines the relaxation of the CO probe, as described by our HRAFM model, the FDBM, with a simplified approach to STM imaging using Chen's approximation, that incorporates both the $\sigma$ and $\pi$ channels, associated with the CO 5$\sigma$ and 2$\pi^*$ orbitals, through which the tunneling current passes via the CO tip. At larger tip-sample distances, the resulting image is a combination of contributions from the $\sigma$ and $\pi$ channels. 
At shorter tip-surface distances, the CO probe relaxation becomes significant, distorting the STM image and enhancing the contrast in the regions near the chemical bonds, similarly to what is observed in HRAFM .

Our results underscore the importance of the substrate effects in the BRSTM image. The substrate induces changes in both  the CO-tip relaxation and the tunneling current.
This could be expected in the  PTCDA /Ag(111) case, where the LUMO gets partially occupied upon adsorption and contributes to the BRSTM images for negative bias (occupied states). However,  these effects are also observable in the TOAT/Cu(111) system, highlighting the need of including the substrate for accurate BRSTM simulations. 
\rs{Our approach provides  an efficient and accurate way to incorporate these effects, as tunneling currents are obtained from the application of Chen's rules directly on the real--space wave functions from the DFT calculation of the full molecule--substrate system.}
The 
\rp{very good} match between simulations and experiments in the observed contrast and its complex evolution with the distance makes our approach a 
\rp{reliable} tool for the interpretation of  BRSTM images, for improving the training of machine learning models for STM structure discovery~\cite{Foster2024STM},  and  for further promoting the use of BRSTM,  an emerging technique that provides intramolecular bond resolution~\cite{Wang2024} using standard STM equipment.
\rp{From a broader perspective, the information gathered from BRSTM images could be even richer, beyond structural determination,  because it involves both forces, that depend on the total charge density, and tunnelling, that is controlled by specific electronic states. Thus, our development should encourage the combined use of HRAFM and BRSTM,  given the complementary information provided by these two techniques.  Recent work on self-assembled CO$_2$ islands on Au(111), both free-standing and confined by 1D metal-organic chains~\cite{CustanceACSNano2024} illustrated the use of simultaneous AFM and STM imaging using CO-functionalized probes to determine the molecular arrangements and to identify the nature of the observed chiral defects.}

\section{Methods}\label{sec:methods}

\subsection{Theoretical calculations}

We performed first-principles, Density Functional Theory (DFT), calculations using the Vienna Ab-initio Simulation Package (VASP) 5.4.4~\cite{kresse1996} with Projected Augmented Wave~\cite{bloch1994} pseudopotentials, the Perdew-Burke-Ernzerhof generalized gradient approximation~\cite{PerdewPRL1996} for the exchange-correlation contribution to the total energy, and the semi-empirical DFT-D3 dispersion correction~\cite{GrimmeJCP2010} for the Van der Waals interaction. The kinetic energy cutoff was set at 400~eV, and the electronic convergence criterium to \(10^{-6}\)~eV.

The initial simulation coordinates for PTCDA/Ag(111) were kindly provided by W. Kami\'nski, see ref.~\onlinecite{Langewisch2012a} for complete details. In short, they include two PTCDA molecules (A and B) in a herringbone arrangement on the (6,1,-3,5) supercell of Ag(111) surface, represented by a slab with three Ag layers. Using the calculation setup above, we first relaxed the system to its ground state  until forces were less than 0.01~eV/\AA{} while fixing the positions of the two bottom Ag layers. The average adsorption distance (\(d_{ads}\), calculated as an average over all the atoms in both PTCDA molecules) was 302~pm, about 20~pm higher than the experimental value.  

To properly simulate changes in the electronic structure, we had to account for the difference between experimental and calculated \(d_{ads}\). We achieve this by artificially lowering the PTCDA molecules to the experimental \(d_{ads}\) and relaxing them again while fixing the z-coordinates of C and H atoms, leaving O and Ag atoms to interact freely. Once we obtained the final adsorption configuration, we extracted the charge density and the Hartree electrostatic potential for the FDBM calculation. For the BRSTM calculation, we recalculated the wavefunctions with a higher, 500~eV, kinetic energy cutoff in order to get smoother partial derivatives.

\subsection{HRAFM and Bond-resolved STM simulations}\label{ssec:methods-brstm}

For the bond-resolved STM simulations with a CO-functionalized tip, we used the FDBM model to calculate the CO position~\cite{ellner2019molecular} and the Tersoff-Hamman and Chen prescriptions to calculate the STM signal from the CO $s$ and \(p_x\) and \(p_y\) channels~\cite{gross2011b}, associated respectively to the  5$\sigma$ and 2$\pi^*$  CO orbitals. 

FDBM first calculates a static potential, $V_{static}$, which includes the short-range Pauli repulsion, the electrostatic interaction, and DFT-D3 dispersion. The short-range interaction is associated with the overlap of tip and sample charge densities, written as
\begin{equation}
        V_{\rm{SR}}(\mathbf{r}_{\rm{probe}}) = V_{0}\int
        \big[\rho_{\rm{probe}}(\mathbf{r}+\mathbf{r}_{\rm{probe}})
        \rho_{\rm{sample}}(\mathbf{r})\big]^{\alpha}d\mathbf{r}.
        \label{eq:fdbmSR}
\end{equation}
Here, $\alpha$ and $V_0$ are fitting parameters that can be optimized to DFT calculations of each particular system, although two universal parameters already provide a very good descrption~\cite{ellner2017,ellner2019}. The parameters used to fit the density overlap integral were \(\alpha = 1.06\) and \(V_0 = 35.16\)~eV\AA{}\(^{3(2\alpha -1)}\). Analogously, the electrostatic interaction comes from the sample's local potential and the charge density of the tip as
\begin{equation}
        V_{\rm{ES}}(\mathbf{r}_{\rm{probe}}) = \int \rho_{\rm{probe}}(\mathbf{r}+\mathbf{r}_{\rm{probe}})
        \phi_{\rm{sample}}(\mathbf{r})d\mathbf{r}. 
        \label{eq:fdbmES}
\end{equation}
$\rho_{\rm{probe}}(\mathbf{r})$, $\rho_{\rm{sample}}(\mathbf{r})$, and $\phi_{\rm{sample}}(\mathbf{r})$   in eqs.~\ref{eq:fdbmSR} and~\ref{eq:fdbmES} are obtained from two separate DFT calculations for tip and sample. 

The last component, the mid- and long-range dispersion interaction $V_{\rm{vdW}}$ is calculated directly with DFT-D3 at each grid point of the static potential. Summing up these three components, we obtain  $V_{static} = V_{\rm{SR}} + V_{\rm{ES}} + V_{\rm{vdW}} $. Forces acting on the tip are calculated in a dense grid as numerical derivatives of the potential~\cite{EVMAppSurfSci2023}.

Finally, the orientation of the CO is determined by placing it in the static potential and minimizing its energy considering that the O atom is effectively the endpoint of a lever with a torsion spring that forces the CO to point straight down 
with a potential energy $\kappa \theta^2/2$, where $\kappa$ is the stiffness of the tip~\cite{EVMAppSurfSci2023}.  This is  equivalent to having a lateral spring with potential energy $kx^2/2$, where, within the small angle approximation, $\kappa=kl^2$,  with the length of the lever $l$ set to 302~pm (the sum of the distances of the C-O and C-metal bonds).  Thus, an stiffness $\kappa =$~0.4 eV/rad\(^2\) is equivalent to $k=$ 0.7 N/m.

For the STM signal, we have implemented a method to calculate the tunneling matrix elements from the sample wavefunctions using Chen's approximation~\cite{Chen1990,Chen1998} for a CO probe~\cite{gross2011b}.
First, we retrieve the eigenvalue of every band and k-point from the VASP calculation with the help of the Atomic Simulation Environment (ASE)~\cite{ASE} software package.
From all the eigenvalues, we select the corresponding bands within the desired range from the Fermi level.
Then, we obtain the real-space values of their wave function in a regular 3D grid with the \textit{vaspwfc} function of the VaspBandUnfolding Python package~\cite{vaspwfc}.
We integrate them directly to obtain the \textit{s}-wave signal ($I_s$)
\begin{equation}
    I_{s}(\mathbf{r}, E) =  \sum_{{\rm{sample}}} \left| \psi_{{\rm{sample}}}(\mathbf{r}) \right|^2 \delta(E_{\rm{sample}} - E)
    \label{eq:swave}
\end{equation}
and we separately take their lateral gradient and then integrate them to obtain the \textit{p}-wave signal:
\begin{equation}
    I_{p}(\mathbf{r}, E) = \sum_{{\rm{sample}}} \left[\left|\frac{\partial \psi_{\rm{sample}}(\mathbf{r})}{\partial x} \right|^2 +
    \left|\frac{\partial \psi_{\rm{sample}}(\mathbf{r})}{\partial y} \right|^2\right] \delta(E_{\rm{sample}} - E).
    \label{eq:pwave}
\end{equation}

Then, both signals were summed with 1:1 weights (as proposed in ref.~\onlinecite{gross2011b}) to obtain the \textit{sp} STM. From this result, we constructed a regular 3D grid where we calculated the interpolated value in the position of the O atom in the relaxed CO-tip for every pixel (\textit{xy}-coordinate) we obtained from the FDBM calculation.

\subsection{Experimental setup}

All experiments were performed in the J\"ulich Quantum Microscope~\cite{Esat2021} at a temperature of 1.2~K and under ultra-high vacuum (UHV) conditions. The Ag(111) surface was prepared in UHV by repeated Ar+ sputtering and heating at 800~K. Sub-monolayer coverages of PTCDA molecules were evaporated onto clean Ag(111) at room temperature. The sample was then flashed at 480~K for 2 min before being cooled to 100~K and transferred to the STM. Carbon monoxide (CO) molecules were naturally abundant on the Ag(111) surface. The STM tip was functionalized with CO molecules according to the procedure described in ref.~\onlinecite{Kichin2011}.

The BRSTM images were acquired with a CO-functionalized STM tip in constant height mode. For this purpose, the tip with the attached CO molecule was placed in the center of a bright PTCDA molecule and the setpoint was adjusted to $I = 20$~nA and $V = -200$~mV. The feedback loop was then turned off and the PTCDA molecules in the layer were scanned at constant height. Current and differential conductance (dI/dV) images were recorded simultaneously. Differential conductance (dI/dV) images were measured using a conventional lock-in technique with an AC modulation amplitude $V_{mod} = 10$~ mV and frequency $f_{mod} = 187$~Hz. The image series at different distances were acquired sequentially by increasing the tip-sample distance in 10 pm steps after each recorded image.


\begin{acknowledgement} 
We thank the financial support from the Spanish Ministry of Science, Innovation and Universities through the Spanish AEI MICIU/AEI/10.13039/501100011033 under project   PID2023-149150OB-I00 and the grant ``Excelencia Maria de Maeztu'' CEX2023-001316-M.
T.E., M.T., R.T. and F.S.T. acknowledge support by the German Federal Ministry of Education and Research through the funding programme ``Quantum technologies: from basic research to market'' under Q-NL (Projects 13N16032 and 13N16046). M.T. acknowledges support from the Heisenberg Program (Grant No. TE 833/2-1)
of the DFG.
Computer time provided by the Spanish Supercomputer Network (RES) at Finisterrae (CESGA, Santiago de Compostela) and Marenostrum (BSC, Barcelona) supercomputers.
\end{acknowledgement}


\section*{Data Availability Statement}

The data and jupyter notebooks necessary to reproduce all the figures in this work are available in the Zenodo repository:  \href{https://zenodo.org/records/15622719}{\texttt{https://zenodo.org/records/15622719}}. \rp{These jupyter notebooks take care of all the practical steps necessary for the simulation of BRSTM images  for the PTCDA and TOAT cases discussed in the paper and can be easily adapted to other systems. These notebooks use the DBSPM package that is freely available in the GitHub repository given below.}


\section*{Code Availability Statement}   

The simulation framework to obtain HRAFM and BRSTM images from two independent DFT calculations for the tip and the sample (molecule + substrate) developed in this work has been implemented in the \texttt{DBSPM} package that is available in the github repository: \href{https://github.com/SPMTH/DBSPM}{\texttt{https://github.com/SPMTH/DBSPM}}. \rp{The current version of DBSPM works only with DFT calculations performed with VASP, but interfaces with other popular DFT codes will be provided in the future.} 

\begin{suppinfo}
High-resolution AFM simulated images of TOAT/Cu(111), including: static potential images, probe deflection maps, and the relaxed HRAFM images.
\end{suppinfo}


\newpage
\bibliography{complete}

\end{document}


\begin{figure}
    \centering
    \includegraphics[width=0.99\textwidth]{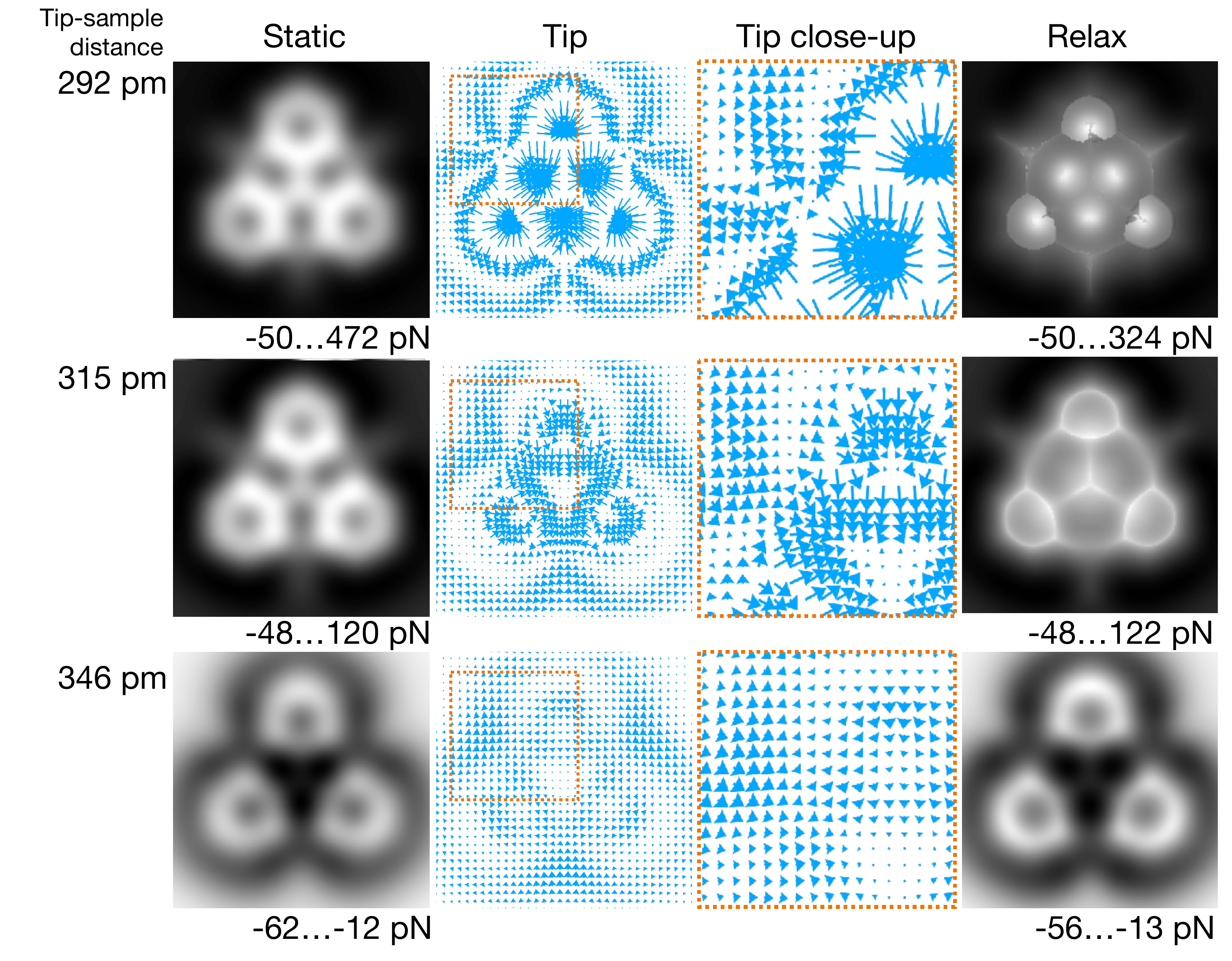}
    \caption{HRAFM images of TOAT/Cu(111) with the static force map $F_{static}$ (left), the deflection of the CO on the tip on the static force map, shown as the shifts (1:1 scale) of the O atom  projected onto the \textit{xy}-plane  (middle left), close-up of the area  delimited by the orange rectangle (middle right), \textit{relaxed} HRAFM images taking into account the CO-tip deflection (right). The gray scale is adjusted to the maximum and minimum value of the force in each image, shown in the labels below each one (positive corresponds to a repulsive force and negative to an attractive one).}
    \label{fig:S1}
\end{figure}